\documentclass[12pt]{article}
\usepackage{epsfig}
\usepackage{psfig}
\def\lsim{\raise0.3ex\hbox{$<$\kern-0.75em\raise-1.1ex\hbox{$\sim$}}}
\def\gsim{\raise0.3ex\hbox{$>$\kern-0.75em\raise-1.1ex\hbox{$\sim$}}}
\makeatletter
\@addtoreset{equation}{section}
\makeatother

\newcommand{\beqn} {\begin{equation}}
\newcommand{\eqn} {\end{equation}}

\newcommand{\slsh}[1] {#1\kern-.43em/}
\newcommand{\real}{{\sf I}\kern-.12em{\sf R}}
\newcommand{\comp}{{\sf I}\kern-.48em{\sf C}}
\newcommand{\nin} {\in\kern-.6em/}

\def\ie{{\sl i.e.~\/}}

\def\MEF{m_{\rm eff}}\def\mef{\ifmmode\MEF\else$\MEF$\fi}

\def\SM{s_{\mu}}\def\xm{\ifmmode\SM\else$\SM$\fi}

\begin{document}
\thispagestyle{empty}
 \hbox{}
 \mbox{} \hfill BI-TP 97/18\\
 \mbox{} \hfill March 1998\\
\begin{center}
{{\large \bf String Tension and Thermodynamics with Tree Level}\\
\medskip
{\large \bf and Tadpole Improved Actions}
} \\
\vspace*{1.0cm}
{\large B. Beinlich, F. Karsch, E. Laermann and A. Peikert} \\
\vspace*{1.0cm}
{\normalsize
{Fakult\"at f\"ur Physik, Universit\"at Bielefeld,
D-33615 Bielefeld, Germany}}
\end{center}
\vspace*{1.0cm}
\centerline{\large ABSTRACT}

\noindent
We calculate the string tension, deconfinement transition 
temperature and bulk
thermodynamic quantities of the $SU(3)$ gauge theory using 
tree level and tadpole improved actions. 
Finite temperature calculations have been performed on lattices
with temporal extent $N_\tau=3$ and 4. Compared to calculations
with the standard Wilson action on this size lattices we observe
a drastic reduction of the cut-off dependence of bulk thermodynamic 
observables at high temperatures. In order to test the influence of
improvement on long-distance observables at $T_c$ we determine the 
ratio $T_c/\sqrt{\sigma}$.
For all actions, including the standard Wilson action, we find results which 
differ only little from each other. 
We do, however, observe an improved asymptotic scaling behaviour 
for the tadpole improved action compared to the Wilson and tree 
level improved actions.

\vskip 20pt
\vfill
\eject

\section{Introduction}

Tree level and tadpole improved actions have been shown to yield a
substantial reduction of cut-off dependences in
the calculation of thermodynamic properties of the $SU(3)$ gauge 
theory \cite{Bei96,Bei97}. In the high (infinite) temperature limit this 
is quite evident already from a perturbative calculation of the pressure
($p$) or energy density ($\epsilon$). In this limit high momentum modes
give the dominant contribution to these observables. An improvement of the
discretization scheme for the Euclidean action at short distances will thus
naturally lead to a better representation of the ideal gas, Stefan-Boltzmann
law on lattices with finite temporal extent $N_\tau$. 
However, even at $T_c$ the improved actions lead to a 
reduced cut-off dependence for some observables. Calculations of the
surface tension or the latent heat of the first order deconfinement 
transition, for instance, show a strong reduction of the cut-off dependence 
on lattices with temporal extent $N_\tau =3$ and 4. In this case it also has 
been found that the remaining cut-off dependence on coarse lattices is weaker
for a tadpole improved action than for a tree-level improved 
action \cite{Bei97}.

At high temperature the variation of $p/T^4$ or $\epsilon /T^4$ with
temperature is small. Both observables only slowly approach the ideal gas 
limit, which is consistent with the expectation that thermodynamics depends
on a running coupling that varies logarithmically with temperature.
The numerical simulations performed in a given discretization scheme at 
finite temperature are therefore particularly sensitive to a correct 
representation of the infinite temperature limit. An accurate determination of
the temperature scale itself is, however, not needed for the observation 
of the dramatic improvement in approaching the continuum 
Stefan-Boltzmann limit at infinite temperature.
This has been utilized in the calculations presented
in \cite{Bei96} where the temperature scale has been fixed using an
effective coupling scheme combined with the asymptotic form of the SU(3)
$\beta$-function \cite{Boy96}. At temperatures
close to the deconfinement transition, however,
thermodynamic observables like $p/T^4$ vary rapidly. A comparison of the 
improvement achieved with different actions in   
this temperature regime requires an accurate determination of the
temperature scale. For some fixed point actions this has been done by
determining the critical temperature on lattices with different
temporal extent \cite{Pap96}. Using calculations of $T_c$ to set the scale
has, however, the disadvantage that $T/T_c$ gets to be known only at a few 
discrete values.  We will use here the string tension to define
a continuous temperature scale, $T/\sqrt{\sigma}$, for thermodynamic
observables. This also is needed for the determination of
thermodynamic observables like the energy density, which require
the knowledge of the $SU(3)~\beta$-function, \ie the derivative of the
bare coupling with respect to the cut-off in the non-asymptotic regime.

In this paper we analyze the heavy quark potential with tree level 
and tadpole improved actions.
From the potential the string tension is extracted in order to define a 
temperature scale, $T/\sqrt{\sigma}$. This in turn is used in a study of the
thermodynamics of the SU(3) gauge theory. In the
next section we discuss our calculations of the heavy quark potential and
the determination of the string tension. The
application of the resulting $\beta$-function for thermodynamic calculations
is presented in Section 3. Section 4 contains our conclusions. 

\section{String Tension}

In our calculations we use two different improved actions for the
SU(3) gauge theory. These actions include in addition to the standard 
$1\times 1$ Wilson loop an additional $1\times 2$ or $2\times 2$ loop,
respectively,
\begin{eqnarray}
S^{(1,2)}  &=&    \sum_{x, \nu > \mu} \biggl( \frac{5}{3}
W^{1,1}_{\mu, \nu}(x)   - {1 \over 6 u_0^2}
W^{1,2}_{\mu, \nu}(x) \biggr)~~,\nonumber \\
S^{(2,2)}  &=&    \sum_{x, \nu > \mu} \biggl( \frac{4}{3}
W^{1,1}_{\mu, \nu}(x)   - {1 \over 48 u_0^4}
W^{2,2}_{\mu, \nu}(x) \biggr)~~.
\label{action}
\end{eqnarray}
Here $u_0$ denotes a tadpole improvement factor which we have chosen
to be defined through the plaquette expectation value \cite{Lep93}, \ie
$u_0 \equiv \bigl(1-\langle W^{1,1}_{\mu, \nu}(x)\rangle  \bigr)^{1/4}$. 
The tree level improved actions
are obtained for $u_0 \equiv 1$. The high temperature ideal gas limit
for these actions has been analyzed previously \cite{Bei96} and
first calculations of the pressure using the action $S^{(2,2)}$ have been
presented there. The analysis of the infinite temperature ideal gas limit
suggests, in fact, that the (1,2)-action is superior to the (2,2)-action.
Although the leading ${\cal O} (a^2)$ corrections are eliminated in both
cases, it turns out that the remaining higher order contributions are
much smaller for the (1,2)-action.
\begin{table}
\begin{center}
\begin{tabular}{|r||r@{(}r|r|}
\hline
\multicolumn{1}{|c||}{$\beta$}&
\multicolumn{2}{c|}{$1-\langle W^{1,1}\rangle$}&
\multicolumn{1}{c|}{$u^2_0$}\\
\hline
\hline
3.80 & 0.463838 &~182~)&0.681057\\
\hline
4.00 & 0.511237 &~275~)&0.715008\\
\hline
4.20 & 0.560142 &~238~)&0.748427\\
\hline
4.50 & 0.613736 &~130~)&0.783413\\
\hline
5.00 & 0.664764 &~114~)&0.815330\\
\hline
6.00 & 0.730540 &~ 74~)&0.854716\\
\hline
\end{tabular}
\end{center} 
\caption{Tadpole improvement factors for the action $S^{(1,2)}$ defined in
Eq.~(\ref{action}).}
\label{tab:tadpole}
\end{table}
In order to calculate the string tension from the long distance part of the 
heavy quark potential we have performed simulations of the $SU(3)$ gauge
theory on lattices of size $16^4$ and $24^4$ at 
several values of the gauge coupling $\beta = 6/g^2$. The heavy quark
potential has been extracted from smeared Wilson 
loops following closely the smearing approach described in \cite{Bal93},
\ie Wilson loops are constructed from smeared links which are obtained
by iterating the replacement process
\beqn
U_\mu (x) \rightarrow U_\mu (x) + \gamma \sum_{\nu \ne \mu} 
U_\nu (x) U_\mu (x+\hat{\nu} ) U^\dagger_\nu (x+\hat{\mu})
\label{smear}
\eqn
several times. Some tests have been performed to find optimal values for
the smearing parameter $\gamma$ for the range of couplings explored here.
The value $\gamma$ varies between
$0.2$ and $0.8$. This allows to achieve a reasonable large overlap with the 
ground state already after about 10 to 15 smearing steps\footnote{Details on
the optimization of our smearing procedure can be found in
Ref.~\cite{Leg97}.}.
The potential at distance $R$ is then determined from the
asymptotic behaviour of  smeared Wilson loops, $W(R , L)$,
\beqn
V(R) = \lim_{L \rightarrow \infty} \ln \biggl(
{W(R , L) \over W(R , L+1)}\biggr)~~ . 
\label{potential}
\eqn
Typically, for each value of the gauge coupling we have analyzed Wilson 
loops on 300 configurations which were separated by 10 updates performed 
with an over-relaxed heat bath algorithm (1 update $\equiv$
4 over-relaxation steps followed by 1 heat bath step). In the case of the
(1,2)-action we also investigated the effect of tadpole improvement. For
this purpose we have first determined self-consistently the
factor $u_0$ at a few values of the gauge coupling. These numbers are given
in Table~\ref{tab:tadpole}. Spline interpolations of these values have then
been used for simulations at other values of the gauge coupling.
Calculations of Wilson loops have
been performed at 10 to 15 values of the gauge coupling. We generally observe
that the ratios of smeared Wilson loops become, within errors, independent
of $L$ already for rather small values, \ie for 
$L > L_{\rm min} \simeq (2-5)$ 
for large $R$. The potential has then been obtained from a weighted average 
of the logarithm of ratios ${W(R , L) / W(R , L+1)}$ with $L > L_{\rm min}$. 
These calculations have been performed on lattices
of size $16^4$, except for the largest values of the gauge coupling where
calculations have been performed on a $24^4$ lattice. 
Similar results have been obtained with the tree level improved action
$S^{(2,2)}$ on lattices of size $24^4$.
The potentials have then been fitted to 
the ansatz 
\beqn
V (R) = V_0 + {\hat{\alpha} \over R} + \hat{\sigma} R ~~,
\label{vfit}
\eqn
for distances $R>R_{\rm min}$ in order to extract the string tension.
For large values of $R$, roughly $Ra>1/4$fm \cite{Alv81}, the 
coefficient of the Coulomb-like term is
expected to be determined by string fluctuations, $\hat{\alpha} = -\pi/12$
\cite{Lue80}. Indeed, for $\beta$ corresponding to a lattice spacing 
smaller than $a \approx 0.13 $fm we find values for $\hat{\alpha}$ 
which are consistent with this (Table~\ref{tab:alpha}).  
\begin{table}[htp]
\catcode`?=\active \def?{\kern\digitwidth}

\begin{center}
\begin{tabular}{|c|c|c|c|l|}\hline
\multicolumn{5}{|c|}{Tree-level improved (1,2)-action}\\
\hline
$N^3_\sigma N_\tau$&$\beta$&$R_{\rm 0.25 fm}$&$R^{\star}_{\rm min}$&
\multicolumn{1}{|c|}{${\hat{\alpha}}$}\\
\hline
$16^4$&4.300& 1.98  & 2.828 & -0.318~(82)\\
\hline                       
$16^4$&4.400& 2.34  & 2.828 & -0.335~(68)\\
\hline                       
$16^4$&4.600& 2.97  & 3.0   & -0.290~(36)\\
\hline                       
$16^4$&4.800& 4.11  & 4.0   & -0.229~(76)\\
\hline                       
$24^4$&5.000& 4.79  & 5.0   & -0.284~(30)\\
\hline
\hline
\multicolumn{5}{|c|}{Tadpole improved (1,2)-action}\\
\hline
$N^3_\sigma N_\tau$&$\beta$&$R_{\rm 0.25 fm}$&$R^{\star}_{\rm min}$
&\multicolumn{1}{|c|}{${\hat{\alpha}}$}\\
\hline
$16^4$&4.600& 1.96 & 2.828 & -0.310~(61)\\
\hline                       
$16^4$&4.800& 2.53 & 2.828 & -0.262~(42)\\
\hline                       
$24^4$&5.100& 3.74 & 4.0   & -0.288~(40)\\
\hline                       
$24^4$&5.250& 4.13 & 4.24  & -0.316~(112)\\
\hline                       
$24^4$&5.400& 4.76 & 5.0   & -0.212~(54)\\
\hline
\hline
\multicolumn{5}{|c|}{Tree level improved (2,2)-action}\\
\hline
$N^3_\sigma N_\tau$&$\beta$&$R_{\rm 0.25 fm}$&$R^{\star}_{\rm min}$
&\multicolumn{1}{|c|}{${\hat{\alpha}}$}\\
\hline
$24^4$&4.600& 1.98 & 2.828 & -0.299~(45)\\
\hline                       
$24^4$&4.800& 2.76 & 2.828 & -0.296~(21)\\
\hline                       
$24^4$&5.000& 3.70 & 4.0   & -0.274~(22)\\
\hline                       
$24^4$&5.200& 4.85 & 5.0   & -0.271~(47)\\
\hline                       
$24^4$&5.400& 6.43 & 5.0   & -0.289~(29)\\
\hline                       
$24^4$&5.600& 8.10 & 5.0   & -0.266~(22)\\
\hline
\end{tabular}
\end{center}
\caption{ Coefficient of the Coulomb-like term obtained from 3 parameter fits 
to the heavy quark potential which have been calculated with tree level 
and tadpole improved (1,2)-actions and tree level improved (2,2) action. 
$R_{\rm 0.25 fm}$ is the distance in lattice units corresponding to a 
physical value of 0.25 fm  and $R^*_{\rm min}$ is the minimal distance 
used in the fits to the potential for the determination of $\hat{\alpha}$. 
The values of alpha and the errors have been calculated in the same way 
as for the string tension according to the formula (\ref{sigma_mean})}
\label{tab:alpha}
\end{table} 
\begin{figure}[ftp]
\begin{center}
   \epsfig{
      file=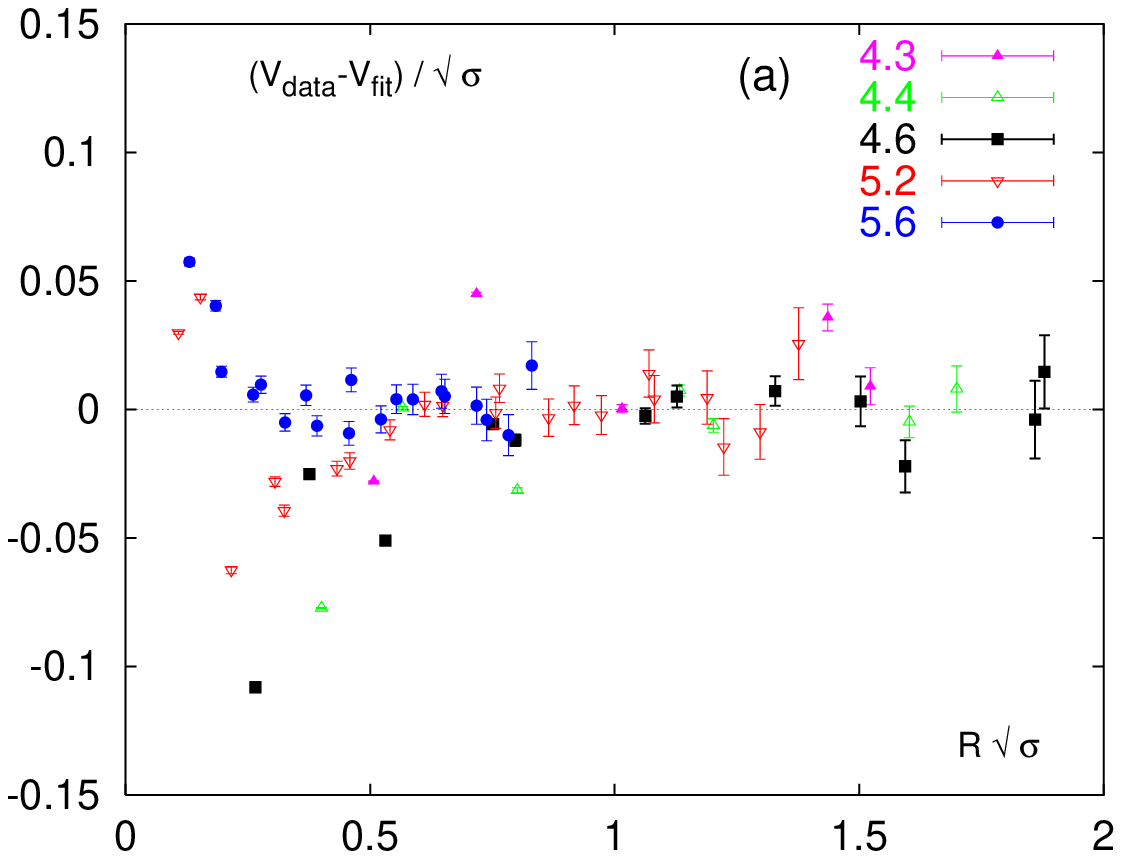,width=130mm}
  \epsfig{
      file=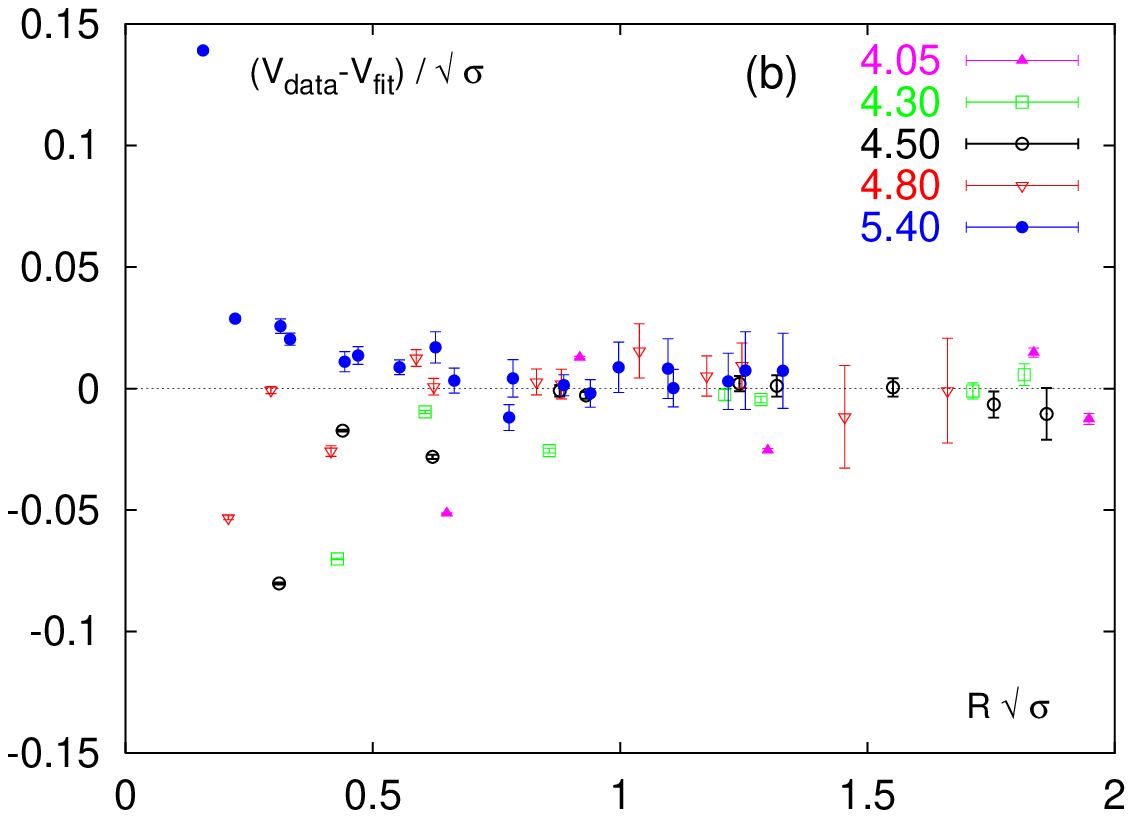,width=130mm}
\end{center}
\caption{
The difference between the calculated potentials and the
best fit obtained with fixed $\hat{\alpha}$ in units of $\sqrt{\sigma}$
versus $R\sqrt{\sigma}$.   Shown are results for the tree level improved 
(2,2)-action (a) and the tadpole improved (1,2)-action (b)
for different values of the gauge coupling. 
Tree level results for the latter action look similar.}
\label{fig:potfig}
\end{figure}
Since we are interested in the long distance behaviour of the
potential we followed the strategy to fix the coefficient of the 
Coulomb-like term for the determination of the string tension. The
minimal value of $R_{\rm min}$ was chosen to be the maximum of 
$Ra = 0.25$fm and $R =~2\sqrt{2}$, so that distortions of rotational
symmetry which are present at small distances are not large. This 
approach was compared to a systematic increase of $R_{\rm min}$  
until stable results have been obtained for $\hat{\sigma}$. Typically 
this was the case for $R^*_{\rm min}\gsim~1/(2\sqrt{\hat{\sigma}})$ 
i.e. $R^*_{\rm min}a\gsim~0.25$fm.
We then have further increased $R_{\rm min}$ and have averaged 
the results for $\hat{\sigma}$ obtained from several
fits with $R_{\rm min} >~R^*_{\rm min}$ in order to minimize remaining
distortion effects resulting from missing rotational invariance of the
potential at these distances. Since the results for $\hat{\sigma}$ at 
distances larger than $R^*_{\rm min}$ agree within errors we calculated 
the mean value according to the fit formula of a constant, but modified 
the error formula by introducing a factor $\sqrt{N}$ which takes into 
account that the different values of $\hat{\sigma}$ are strongly correlated.
\begin{eqnarray}\label{sigma_mean}
\hat{\sigma}&=&\left(\sum_{\hat{R}_{\rm min}\ge \hat{R}^\star_{\rm min}}
{\hat{\sigma}(\hat{R}_{\rm min})\over {(\Delta\hat{\sigma}(\hat{R}_
{\rm min}))^2}}\right) \left(\sum_{\hat{R}_{\rm min}\ge \hat{R}^
\star_{\rm min}}
{1\over (\Delta\hat{\sigma}(\hat{R}_{\rm min}))^2}\right)^{-1}\nonumber\\
\Delta\hat{\sigma}&=&\left({1\over N}\sum_{\hat{R}_{\rm min}\ge 
\hat{R}^\star_{\rm min}}
{1\over (\Delta\hat{\sigma}(\hat{R}_{\rm min}))^2}\right)^{-{1\over2}}~~.
\end{eqnarray}
The resulting values for the string tension for the different actions 
are given in Table~\ref{tab:string} and in Table~\ref{tab:str}.
Also given there are the minimal values $L_{\rm min}$ and $R^*_{\rm min}$ 
used to extract and fit the potential for a given value of
$\beta$. In the last 
column we show the values of $\hat{\sigma}$ obtained from the 3
parameter fit of the potential (Eq.~\ref{vfit}). Except for
$\beta=4.400$ 
for the tree level
improved (1,2) and $\beta=4.500$ for the tree level improved (2,2) action 
all values of the  string tensions obtained from 2 parameter fits agree 
within errors with the ones from 3 parameter fits. The $\chi^2$ of the 
fits is comparable.

In Figure~\ref{fig:potfig} we show the difference between the calculated 
potential and the best fit obtained with $\hat{\alpha} = -\pi/12$.
Results are shown for the tree level improved (2,2)-action and the tadpole 
improved (1,2)-action. All fits have been performed for distances 
$R\sqrt{\hat{\sigma}} \gsim~0.5$. We note that the agreement between the 
numerical data and the fit generally remains good at distances smaller than 
the fitting interval. For smaller values of $\beta$ we observe,
however, a scattering of the data for $R\sqrt{\hat{\sigma}} <~0.5$ due to the 
lack of rotational invariance. This effect becomes smaller for
larger values of $\beta$, \ie closer to the continuum limit. For 
large values of $\beta$ the potential can also be obtained at
shorter physical distances. Here we clearly 
observe that the potential at small distances would prefer a
larger coupling for the Coulomb term than used in our fit. This is
reflected in the increase of 
$V_{\rm data}-V_{\rm fit}$ at short distances. 
\begin{table}[htp]
\begin{center}
\begin{tabular}{|c|c|c|c|l|l|}
\hline
\multicolumn{6}{|c|}{Tree-level improved (1,2)-action}\\
\hline
$N^3_\sigma N_\tau$&$\beta$&$L_{\rm min}$&$R^{\star}_{\rm
  min}$&\multicolumn{1}{|c|}{${\hat{\sigma}_{\rm 2par}}$}&
\multicolumn{1}{|c|}{${\hat{\sigma}_{\rm 3par}}$}\\
\hline
\hline
$16^4$& 3.850 & 1 & 2.828 & 0.35173~(328) & 0.35789~(1327)\\
\hline                              
$16^4$& 4.000 & 2 & 2.828 & 0.19923~(162) & 0.19748~(836)\\
\hline                              
$16^4$& 4.040 & 2 & 2.828 & 0.17542~(104) & 0.17288~(457)\\
\hline                              
$16^4$& 4.100 & 3 & 2.828 & 0.14234~(165) & 0.13871~(842)\\
\hline                              
$16^4$& 4.150 & 3 & 2.828 & 0.11423~(131) & 0.11079~(586)\\
\hline                                    
$16^4$& 4.300 & 4 & 2.828 & 0.07265~(107) & 0.06987~(489)\\
\hline                                    
$16^4$& 4.400 & 4 & 2.828 & 0.05412~(51)  & 0.05055~(182)\\ 
\hline                                     
$16^4$& 4.600 & 4 & 3.0   & 0.03201~(43)  & 0.03105~(182)\\ 
\hline                                     
$16^4$& 4.800 & 5 & 4.0   & 0.01634~(48)  & 0.01734~(257)\\ 
\hline                                     
$24^4$& 5.000 & 6 & 5.0   & 0.01188~(18)  & 0.01104~(84)\\ 
\hline
\hline
\multicolumn{6}{|c|}{Tadpole improved (1,2)-action}\\
\hline
$N^3_\sigma N_\tau$&$\beta$&$L_{\rm min}$&$R^{\star}_{\rm
  min}$&\multicolumn{1}{|c|}{${\hat{\sigma}_{\rm 2par}}$}&
\multicolumn{1}{|c|}{${\hat{\sigma}_{\rm 3par}}$}\\
\hline
\hline
$16^4$&4.050& 1 & 2.828 & 0.42159~(217) & 0.42216~(1462)\\
\hline           
$16^4$&4.150& 2 & 2.828 & 0.29927~(340) & 0.32253~(3625)\\
\hline           
$16^4$&4.185& 2 & 2.828 & 0.27153~(216) & 0.25530~(1645)\\
\hline             
$16^4$&4.200& 2 & 2.828 & 0.26205~(198) & 0.25367~(1124)\\
\hline                   
$16^4$&4.250& 2 & 2.828 & 0.21829~(176) & 0.21355~(893)\\ 
\hline                                    
$16^4$&4.300& 2 & 2.828 & 0.18343~(103) & 0.17852~(496)\\
\hline                                    
$16^4$&4.350& 3 & 2.828 & 0.15416~(238) & 0.15947~(1265)\\   
\hline           
$16^4$&4.400& 3 & 2.828 & 0.12994~(154) & 0.12581~(754)\\
\hline           
$16^4$&4.500& 3 & 2.828 & 0.09633~(53)  & 0.09536~(252)\\
\hline           
$16^4$&4.600& 4 & 2.828 & 0.07154~(90)  & 0.06839~(366)\\
\hline           
$16^4$&4.800& 4 & 2.828 & 0.04315~(55)  & 0.04296~(235)\\
\hline           
$24^4$&5.100& 6 & 4.0   & 0.01967~(22)  & 0.01906~(87)\\
\hline           
$24^4$&5.250& 6 & 4.24  & 0.01604~(53)  & 0.01513~(233)\\
\hline           
$24^4$&5.400& 6 & 5.0   & 0.01227~(20)  & 0.01308~(97)\\
\hline
\end{tabular}
\end{center}
\caption{ String tension obtained from 2 and 3 parameter fits to the heavy 
quark  potential which have been calculated with tree level and 
tadpole improved  (1,2)-actions. $L_{\rm min}$ is the minimal extent 
of smeared Wilson loop $W(R,L)$ used to extract the potential 
at distance $R$. $R^*_{\rm min}$ is the minimal distance used in fits 
to the potential for the determination of the string tension. }
\label{tab:string}
\end{table} 
\begin{table}[htp]
\begin{center}
\begin{tabular}{|c|c|c|l|l|}
\hline
\multicolumn{5}{|c|}{Tree-level improved (2,2)-action}\\
\hline
$\beta$&$L_{\rm min}$&$R^{\star}_{\rm
  min}$&\multicolumn{1}{c}{${\hat{\sigma}_{\rm
      2par}}$}&\multicolumn{1}{|c|}{${\hat{\sigma}_{\rm 3par}}$}\\
\hline
\hline
4.300&3& 2.828 & 0.24547~(981) & 0.31077~(7679)\\
\hline
4.380&3& 2.828 & 0.17401~(278) & 0.15589~(1534)\\
\hline
4.396&3& 2.828 & 0.16334~(228) & 0.16059~(1366)\\
\hline
4.400&3& 2.828 & 0.16052~(207) & 0.16778~(1474)\\
\hline
4.406&3& 2.828 & 0.15603~(186) & 0.14942~(1257)\\
\hline
4.430&3& 2.828 & 0.13882~(141) & 0.13238~(900)\\
\hline
4.450&4& 2.828 & 0.12787~(102) & 0.12698~(1949)\\
\hline
4.500&4& 2.828 & 0.10422~(53)  & 0.08755~(867)\\
\hline
4.550&4& 2.828 & 0.08636~(83)  & 0.08431~(461)\\
\hline
4.600&4& 2.828 & 0.07060~(48)  & 0.06957~(236)\\
\hline
4.800&4& 2.828 & 0.03606~(13)  & 0.03536~(57)\\
\hline
5.000&4& 4.0   & 0.02046~(13)  & 0.02021~(48)\\
\hline
5.200&5& 5.0   & 0.01169~(16)  & 0.01166~(76)\\
\hline
5.400&5& 7.0(5.0) & 0.00669~(14) & 0.00642~(46)\\
\hline
5.600&5& 8.0(5.0) & 0.00426~(14) & 0.00425~(35)\\
\hline
\end{tabular}
\end{center}
\caption{String tension obtained from 2 and 3 parameter fits to the 
heavy quark potential which have been calculated with the tree level improved
(2,2)-action on lattices of size $24^4$. $L_{\rm min}$ and $R^*_{\rm min}$
are explained in Table \ref{tab:string}. For the last two $\beta$
values we used a smaller
value of $R^*_{\rm min}=5.00$ in the 3 parameter fit to obtain stable 
results. }
\label{tab:str}
\end{table}
\subsection{Deviations from asymptotic scaling}  

Improved actions depend on several couplings for the different
Wilson loops appearing in the action. Through the specific choice
of these couplings a particular trajectory in a multi-dimensional
parameter space is defined on which the continuum limit is
approached.  Although the improvement scheme for the actions does
not aim at improving the approach to asymptotic scaling we may
test in how far the scaling behaviour is modified through the
specific choice of a trajectory.  The results for the string
tension given in Table~\ref{tab:string} and \ref{tab:str} can be
used to analyze the relation between the bare gauge coupling and
the lattice cut-off, $\beta(a) \equiv 6/g^2(a)$.  In the case of
the Wilson action this has been analyzed in quite some detail
\cite{Ake93}.
\begin{figure}[ftp]
\centerline{
   \epsfig{
      file=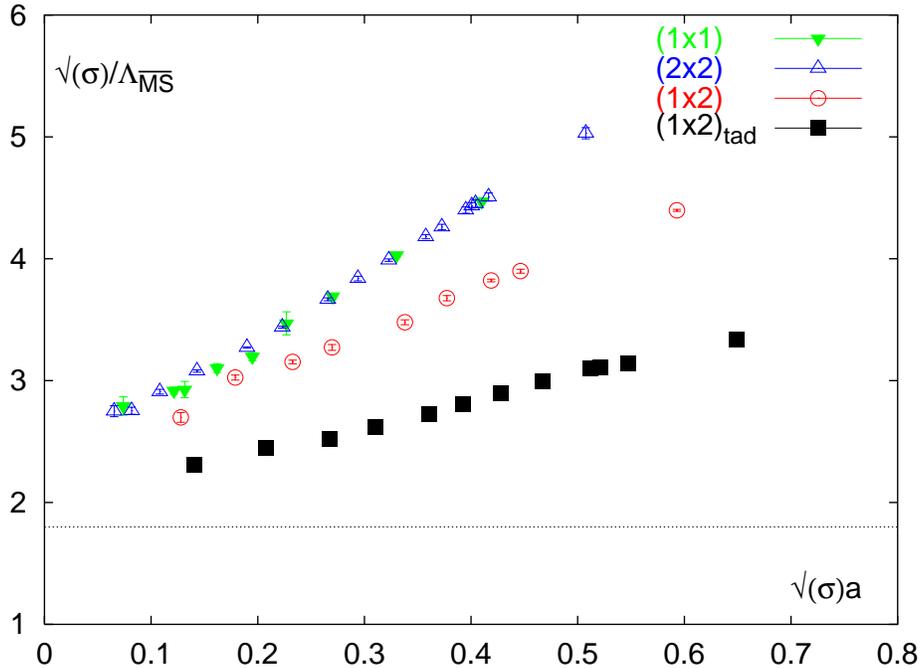,width=130mm}}
\caption{
$\sqrt{\sigma}/\Lambda_{\rm \overline{MS}}$ versus $\sqrt{\sigma}a$ for 
various actions.
Shown are results for $\sqrt{\sigma}a/R(g^2)$, normalized to $\Lambda_{\rm
\overline{MS}}$. Here $R(g^2)$ denotes the 2-loop $\beta$-function. In the
case of the (2,2)-action the ratio of 
$\Lambda^{(2,2)}_{\rm tree} /\Lambda_{\rm \overline{MS}}$ 
was not known to us and we have, therefore, used an arbitrary
normalization factor of 0.15. The horizontal line indicates the value of the
continuum extrapolation of $\sqrt{\sigma}/\Lambda_{\rm \overline{MS}}$ for the
Wilson action taken from Ref.~[6].}
\label{fig:sigmascaling}
\end{figure}
In a most straightforward way the deviations from asymptotic
scaling become visible when one divides the results obtained for
$\sqrt{\sigma}a$ by the universal 2-loop form of the
$\beta$-function. This yields $\sqrt{\sigma}/\Lambda_L$. The
perturbative expansion of the tadpole improved (1,2)-action
coincides up to ${\cal O} (g^2)$ with that of the tree level
improved (1,2)-action with a modified gauge coupling
$\tilde{\beta} \equiv \beta (1+\alpha_2 g^2 /3)$, where $\alpha_2
= -0.366263 (N^2-1)/4N$ denotes the ${\cal O} (g^2)$ expansion
coefficient for the plaquette expectation value calculated with
the tree level improved action \cite{Bei96}. From this we obtain
the ratio
of $\Lambda$-parameters for tree level and tadpole improved (1,2)-actions,
\begin{equation}
{\Lambda_{\rm tad}^{(1,2)} \over \Lambda_{\rm tree}^{(1,2)}} = 
e^{\alpha_2 /6b_0} \quad ,
\label{ratios}
\end{equation}
with $b_0 = 11N/48\pi^2$. Using also the ratios of lattice 
$\Lambda$-parameters to $\Lambda_{\rm \overline{MS}}$ \cite{Has80,Ber83} we 
obtain $\sqrt{\sigma}/\Lambda_{\rm \overline{MS}}$ for different actions. 
This is shown in Figure~\ref{fig:sigmascaling}. The numbers are plotted
versus $\sqrt{\sigma}a$ in order to be able to compare results
at the same value of the cut-off. The slope of these curves
is an indication for the deviations from asymptotic scaling in the regime 
of couplings investigated.
We note that the tree level improved (2,2)-action shows
similar scaling violations as the Wilson action while these are 
reduced for the tree level and even more for the tadpole improved 
(1,2)-actions. A similar behaviour is also obtained from calculations of 
the critical temperature with the tree level improved 
(1,2) action \cite{Cel94}.

\section{Thermodynamics}

\subsection{$T_c / \sqrt{\sigma}$}
      
The string tension calculated in the previous section can be used
to fix the temperature scale $T/\sqrt{\sigma} = 1/(N_\tau
\sqrt{\hat{\sigma}})$. In particular, we have calculated the ratio
$T_c / \sqrt{\sigma}$ by determining the critical couplings for
the different tree level and tadpole improved actions on lattices
with temporal extent $N_\tau = 3$ and 4. Pseudo-critical couplings
on finite spatial lattices have been determined from the location
of the peak in the Polyakov loop susceptibility \cite{Boy96}. For
the tree level and tadpole improved (1,2)-actions we, furthermore,
have performed a detailed study of the finite volume dependence of
the critical couplings on lattices with temporal extent $N_\tau
=4$ and $N_\sigma=16$, 24 and 32 \cite{Bei97}.  In these cases the
critical couplings have been extrapolated to the
infinite volume limit using the ansatz
\beqn \beta_c(N_\tau, N_\sigma) = \beta_c (N_\tau,\infty) - h
\biggl( {N_\tau \over N_\sigma} \biggr)^3
\label{extrapolation}
\eqn
which is appropriate for first order phase transitions. We note that in the
continuum limit the difference
$\beta_c(N_\tau,\infty)-\beta_c(N_\tau,N_\sigma)$ is proportional to the
finite volume shift in the critical temperature, i.e.
$(T_{c,\infty}-T_{c,V})/T_{c,\infty}$. 
In this limit Eq.~(\ref{extrapolation}) thus
characterizes a physical property of QCD in a finite volume, the finite
size dependence of peaks in susceptibilities which develop into a
singularity in the infinite volume limit. Eq.~(\ref{extrapolation}) then reads
\begin{equation}
{T_{c,\infty}-T_{c,V} \over  T_{c,\infty} } = {4{\pi}^2 \over 33}~{h \over VT^3} 
\end{equation}
The parameter $h$ thus is independent of the lattice action up to
finite cut-off corrections. Given the current uncertainties in the
numerical values of $h$ these finite cut-off corrections can, however,
not be disentangled from the statistical errors.\\
We find for the tree level and tadpole improved (1,2)-actions proportionality 
factors $h$ which are 
consistent with each other as well as with earlier results obtained with 
other actions,
\beqn
h = \cases{ 
0.082~(32) & (1,1) Wilson action \cite{Iwa92}, $N_\tau=4$ \cr
0.072~(77) & (1,1) Wilson action \cite{Iwa92}, $N_\tau=6$ \cr
0.101~(34) & (1,2) tree level action, $N_\tau=4$ \cr
0.068~(45) & (1,2) tadpole action, $N_\tau=4$ \cr
0.122~(54) & RG action \cite{Iwa97}, $N_\tau=3$ \cr
0.133~(63) & RG action \cite{Iwa97}, $N_\tau=4$ \cr}
\label{proportional}
\eqn
The calculated critical couplings as well as the extrapolations to the 
infinite volume limit are summarized in Table~\ref{tab:critical}
for the different actions. We note that our results for the tree level 
improved (1,2)-action are consistent with those of Ref.~\cite{Cel94}. In 
Table~\ref{tab:critical} we also give results from \cite{Cel94} for larger
values of $N_\tau$ and use the ansatz of Eq.~(\ref{extrapolation}) to 
extrapolate to the critical couplings on an infinite lattice. For the 
constant $h$ we use in these cases the weighted average of the values 
given in Eq.~(\ref{proportional}), \ie $h= 0.093~(18)$. 
\begin{table}
\begin{center}
\begin{tabular}[t]{|c|c|l|l|} 
\hline
\multicolumn{4}{|c|}{Tree level improved (1,2)-action}\\
\hline
$N^3_\sigma$&$N_\tau$& \multicolumn{1}{|c|}{$\beta_c$}&{$T_c/\sqrt{\sigma}$}\\
\hline \hline
$12^3$ & 3 &  ~~3.9079~(6)&~\\
\hline
$(\infty)^3 $& 3&  ~~3.9094~(6)(3)&0.630~(5)\\
\hline\hline
$32^3$ & 4 &  ~~4.0729~(3)&~\\
\hline
$(\infty)^3 $& 4 & ~~4.0730~(3)&0.636~(4)\\
\hline \hline
$20^3$ & 5 & ~~4.19963~(14)&~\\
\hline
$(\infty)^3 $& 5 & ~~4.20108~(14)(28)&0.631~(5)\\
\hline \hline
$24^3$ & 6 & ~~4.31466~(24)&~\\
\hline
$(\infty)^3 $& 6 & ~~4.31611~(24)(28)&0.632~(5)\\
\hline\hline
\multicolumn{4}{|c|}{Tree level improved (2,2)-action}\\
\hline
$N^3_\sigma$&$N_\tau$& \multicolumn{1}{|c|}{$\beta_c$}&{$T_c/\sqrt{\sigma}$}\\
\hline \hline
$24^3$ & 4 &  ~~4.3995~(2)&~\\
\hline
$(\infty)^3 $& 4&  ~~4.3999~(2)(1)&0.625~(4)\\
\hline\hline
\multicolumn{4}{|c|}{Tadpole improved (1,2)-action}\\
\hline
$N^3_\sigma$&$N_\tau$& \multicolumn{1}{|c|}{$\beta_c$}&{$T_c/\sqrt{\sigma}$}\\
\hline \hline
$12^3$ & 3 & ~~4.1868~(4)&~\\
\hline
$(\infty)^3 $& 3 & ~~4.1882~(4)(3)&0.643~(3)\\
\hline\hline
$32^3$ & 4 &  ~~4.3522~(4)&~\\
\hline
$(\infty)^3$ & 4 & ~~4.3523~(4)&0.639~(6)\\
\hline
\end{tabular}
\end{center}
\caption{ Critical couplings for the tree level and tadpole improved 
  actions. In each case we give the result for the largest spatial
  lattice on which simulations have been performed and the
  infinite volume extrapolation. Details on the determination of
  critical couplings for the (1,2)-actions for $N_\tau = 3$ and 4
  on lattices with different spatial extent are given in [2].
  The finite lattice results for $N_\tau = 5$ and 6
  are taken from [13]. Infinite volume extrapolations are
  based on Eq.~(\ref{extrapolation}) with $h=0.093$, except for
  the case of the $N_\tau=4$ (1,2)-actions where a detailed finite
  volume scaling analysis has been performed in [2]. The
  second error on $\beta_c$ in the infinite volume limit 
 is the systematic error due to the error on
  $h$. In the last column we also give the ratio
  $T_c/\sqrt{\sigma}$. The errors on $T_c/\sqrt{\sigma}$ due to the
  error on $\beta_c$ and on $\sqrt{\hat{\sigma}}$ have been added. }
\label{tab:critical}
\end{table}

\begin{figure}[ftp]
\begin{center}
      \epsfig{file=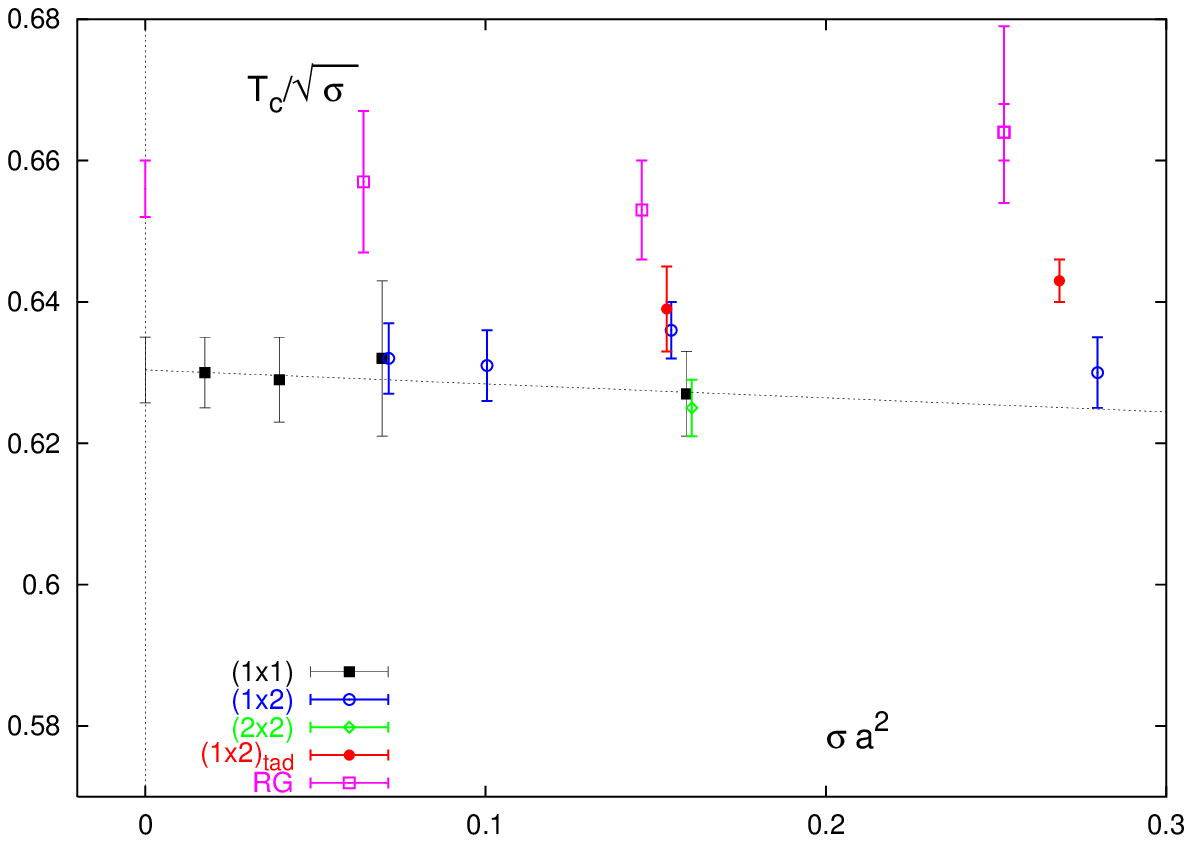,width=140mm}
\end{center}
\caption{The critical temperature in units of the square root of the string 
tension  for various actions versus the square of the cut-off. The 
$N_{\tau}=6$ point of the tree level (1,2) improved action has
been slightly shifted 
to make it distinguishable from the Wilson action point.}
\label{fig:tcsigma}
\end{figure}

In order to extract the critical temperature in units of the square root
of the string tension we determine $\hat{\sigma}$ at $\beta_c(N_\tau,
\infty)$ from an interpolation with an exponential ansatz, 
$\hat{\sigma} = A\exp(-\beta /2Nb_0+f(\beta))$, where $f(\beta)$ is a
third order polynomial in $\beta^{-1}$. The results 
for $T_c/\sqrt{\sigma}$ are shown in 
Table~\ref{tab:critical} and in Figure~\ref{fig:tcsigma}.\\  
The Wilson action $\beta_c(N_{\tau,\infty})$ for $N_{\tau}=$4
and 6 are taken from \cite{Iwa92}. For $N_{\tau}=8$ 
and $N_{\tau}=12$ we performed the infinite volume extrapolation based on 
Eq.~(\ref{extrapolation}) with $h=0.093$ by using $\beta_c$ from 
\cite{Boy96}.  Recently there has been a new 
measurement of the string tension for the Wilson gauge action 
by Edwards et al. \cite{Edw97}. We used their ``best''
parameterization of the 
string tension to determine $\sqrt{\hat{\sigma}}$ at $\beta_c$.
The results are displayed in Table~\ref{tab:critwils} and 
Figure~\ref{fig:tcsigma}. The strong cut-off effect so far reported for 
$T_c/\sqrt{\sigma}$ is no longer visible in these new results 
\footnote{Previous estimates of $T_c / \sqrt{\sigma}$ for the Wilson 
action at low $\beta$ were based on preliminary results for the string 
tension \cite{Bor91} which propagated through the literature. Published 
values \cite{Bor94} and recent high statistics results \cite{Edw97} for 
the string tension are lower than the preliminary ones and thus lead to 
the increase in $T_c / \sqrt{\sigma}$ at $N_{\tau}=$~4 and 6.}.
For the continuum extrapolation we find the following value
\beqn 
{T_c \over \sqrt{\sigma}} = 0.630 \pm 0.005~~,
\label{ratio}
\eqn
which is consistent with all ratios ${T_c / \sqrt{\sigma}}$ extracted by 
us with improved actions on $N_\tau \ge 4$ lattices 
(see Table~\ref{tab:critical}).
\begin{table}
\begin{center}
\begin{tabular}[t]{|c|c|l|l|}
\hline
\multicolumn{4}{|c|}{Wilson action}\\
\hline
$N^3_\sigma$&$N_\tau$& \multicolumn{1}{|c|}{$\beta_c$}&{$T_c/\sqrt{\sigma}$}\\
\hline \hline
$(\infty)^3 $ & 4 &  ~~5.6925~(2)& ~ 0.627~(6)\\
\hline
$(\infty)^3 $ & 6 &  ~~5.8941~(5)& ~ 0.632~(11)\\
\hline
$(\infty)^3 $ & 8 & ~~6.0624~(9)~(3)& ~ 0.629~(6)\\
\hline
$(\infty)^3 $ & 12 & ~~6.3380~(13)~(10)& ~ 0.630~(5)\\
\hline
\end{tabular}
\end{center}
\caption{ Critical couplings
for the Wilson gauge action. We give the results for the infinite volume 
extrapolations for $N_{\tau}=4$ and 6 from [15] and for 
$N_{\tau}=8 $ and 12 based on Eq.~(\ref{extrapolation}) with
$h=0.093$, where the values for $\beta_c$ on finite lattices are taken 
from [3]. The second error given is the systematic error due to the 
uncertainty in the extrapolation. In the last column we give the ratio 
$T_c/\sqrt{\sigma}$ using the string tension parameterization given in 
[14].}
\label{tab:critwils}
\end{table}
In Figure~\ref{fig:tcsigma} and Table~\ref{tab:tc} we give the results 
obtained from calculations with different actions. This also includes 
results obtained with a renormalization group improved action \cite{Iwa97}.
The latter does lead to a slightly larger value for $T_c / \sqrt{\sigma}$.
However, also in this case the cut-off dependence is comparable to the other 
actions. This suggests that the difference between the
results obtained with the RG-improved action in 
Ref.~\cite{Iwa97} and the results presented here is mainly due to
differences in the analysis of the heavy quark potential rather than due
to differences in the improvement scheme. In general we find that the
cut-off dependence in the ratio $T_c/\sqrt{\sigma}$ is quite small for all
actions. We also note that the results 
on $N_\tau=3$ lattices are in good agreement with those obtained on larger
lattices. This is
quite different from the surface tension analysis \cite{Bei97}. It may, 
however, indicate that the latter is more sensitive to high momentum modes 
than the ratio $T_c/ \sqrt{\sigma}$, which is also reflected in the fact 
that the Wilson action does not show large finite cut-off effects in 
$T_c / \sqrt{\sigma}$ whereas there are large finite cut-off dependences 
in the surface tension.
%
%
\begin{table}[hbt]
\begin{center}
\begin{tabular}{|l|l|l|}
\hline
action &
$\beta_c$&
$T_c/\sqrt{\sigma} $\\ \hline
standard Wilson  &5.69254~(24)&$ 0.627~(6)$\\
(2,2) (tree level improved) &4.3999~(3)&$ 0.625~(4) $\\
(1,2) (tree level improved) &4.0730~(3)&$ 0.636~(4) $\\
(1,2) (tadpole improved) &4.3523~(4)&$ 0.639~(6) $\\
(1,2) (RG improved) &2.2879~(11)&$ 0.653~(6)(1) $\\ 
\hline
\end{tabular}
\end{center}
\caption{Critical temperature in units of $\sqrt{\sigma}$ on lattices 
with temporal extent $N_\tau=4$. Infinite volume extrapolations for the
critical couplings have been performed in all cases. Further details on the
data for the RG-improved action can be found in [18].}
\label{tab:tc}
\end{table}

\subsection{Pressure of the SU(3) gauge theory}

\begin{figure}[ftp]
\begin{center}
\vspace*{-0.4cm}
   \epsfig{
      file=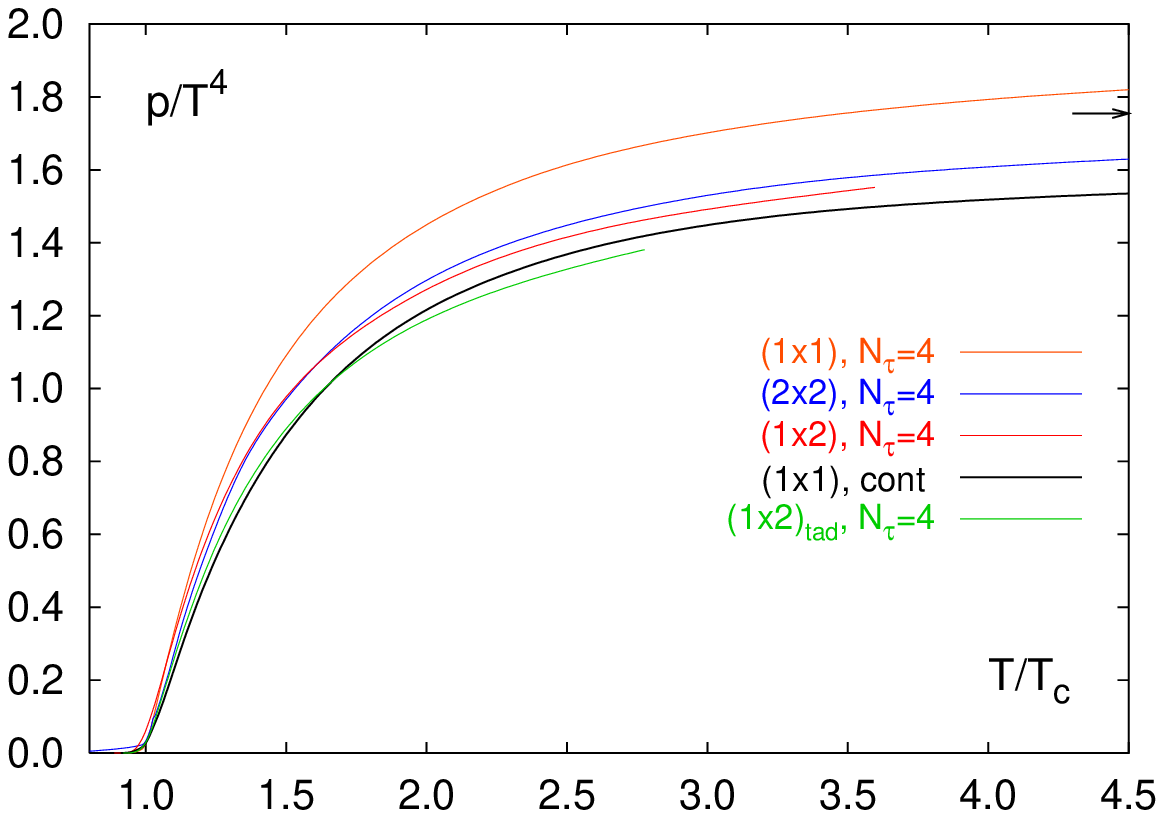,width=130mm}
%
\vspace*{-0.4cm}
  \epsfig{
      file=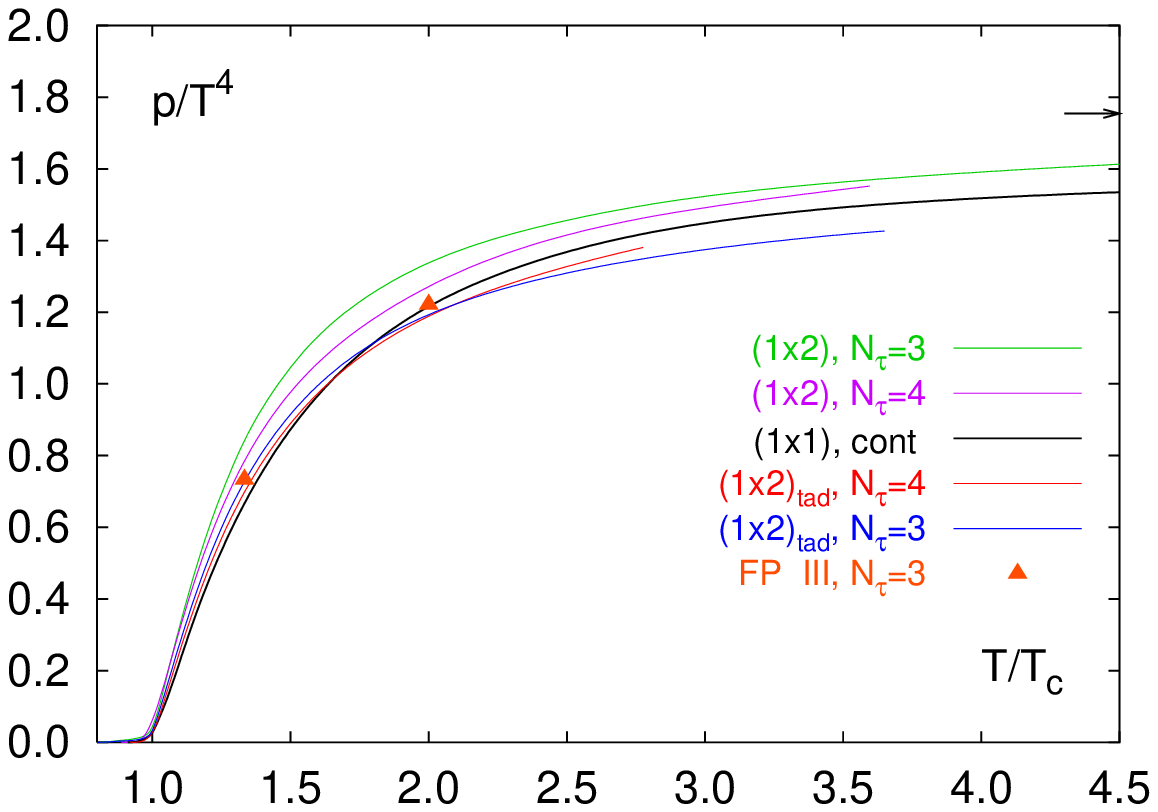,width=130mm}
\end{center}
\caption{Pressure of the SU(3) gauge theory calculated with
the Wilson action and different improved actions on $N_\tau=4$ lattices
(upper figure). The lower figure shows a comparison of calculations with
tree level and tadpole improved actions on $N_\tau=3$ and 4 lattices.
Also shown there are results from a calculation with a fixed point action
(triangles).
The arrows indicate the ideal gas result in the continuum limit.
For comparison we also show the continuum extrapolation obtained from
calculations with the Wilson action [3].
}
\label{fig:pressure}
\end{figure}

The temperature dependence of the pressure can be obtained from
an integration of action densities, $\langle S \rangle$, 
calculated at zero and non-zero
temperature, respectively\footnote{We refer to
Refs.~\cite{Boy96,Eng95} for more details on the formalism.},
\beqn
{p\over T^4}\Big\vert_{\beta_0}^{\beta}
=~N_\tau^4 \int_{\beta_0}^{\beta}
{\rm d}\beta'  (\langle \tilde{S} \rangle_0 - \langle \tilde{S} \rangle_T )~~.
\label{freelat}
\eqn
Here $\beta_0$ is a value of the coupling constant below the phase
transition
point at which the pressure can safely be approximated by zero.
The subscripts 0 and T refer to calculations of the action expectation
values on the zero temperature and non-zero temperature
lattices, respectively. 
We also note that in the case of tadpole improved actions
the action itself depends implicitly on the gauge coupling through the
tadpole factor $u_0(\beta)$. This has to
be taken into account in the calculation of derivatives with respect 
to $\beta$.
We therefore have introduced in Eq.~(\ref{freelat}) the quantity $\tilde{S}$ 
defined as 
\beqn
\tilde{S} = S - \beta~{{\rm d} S \over {\rm d} \beta}~~. 
\label{stilde}
\eqn
Using the string tension values given in Table~\ref{tab:string}
and \ref{tab:str} and
normalizing these to the string tension at $\beta_c$, we obtain the 
temperature in units of $T_c$. The temperature at intermediate values of the
coupling has been obtained from an interpolation. 
With this we can reanalyze also the results 
for the pressure obtained for the tree level improved
(2,2)-action given in \cite{Bei96}. It simply amounts to a modification of 
the temperature scale, i.e. the ordinate of
Figure~3 in \cite{Bei96}. While this has practically no consequences at high
temperature, it leads to visible shifts close to $T_c$. In 
Figure~\ref{fig:pressure} we show the results of calculations with improved
actions on $16^3\times 4$ lattices. These are compared to the continuum
extrapolation obtained from calculations with the Wilson action on lattices
with temporal extent $N_\tau=4$, 6 and 8 \cite{Boy96}. We find that all 
improved action calculations are close to the continuum extrapolation, while
the standard Wilson calculation on a $N_\tau=4$ lattice clearly deviates
substantially. 

We furthermore have calculated the pressure for the tree level and tadpole
improved (1,2)-actions on a lattice of size $12^3 \times 3$. These
results are shown in Figure~\ref{fig:pressure}b.
Here we also show the result of a calculation using a fixed point action
for $N_\tau=3$ \cite{Pap96}. These are in good agreement with the continuum
extrapolation obtained from the Wilson action and seem to be closer to the
result obtained with the tadpole improved action than the tree level result.
For temperatures in the range $(2-3)T_c$ the tadpole action yields 
about 10\% smaller values for the pressure than the tree level action, 
although both approach the same 
infinite temperature limit. We also note that we do not observe any 
significant 
cut-off dependence when comparing calculations on lattices with temporal 
extent $N_\tau =3$ and 4 despite the fact that the cut-off dependence in the
infinite temperature limit leads to about 15\% differences in the
Stefan-Boltzmann limit. This is, to some extent, in accordance with the
analysis of the Wilson action, where we noted already that the cut-off
distortion in this temperature range is only half as large as expected on the 
basis of calculations for the ideal gas limit \cite{Boy96}. It seems that 
these are further reduced in calculations with improved actions. 
\begin{figure}[ftp]
\begin{center}
   \epsfig{
      file=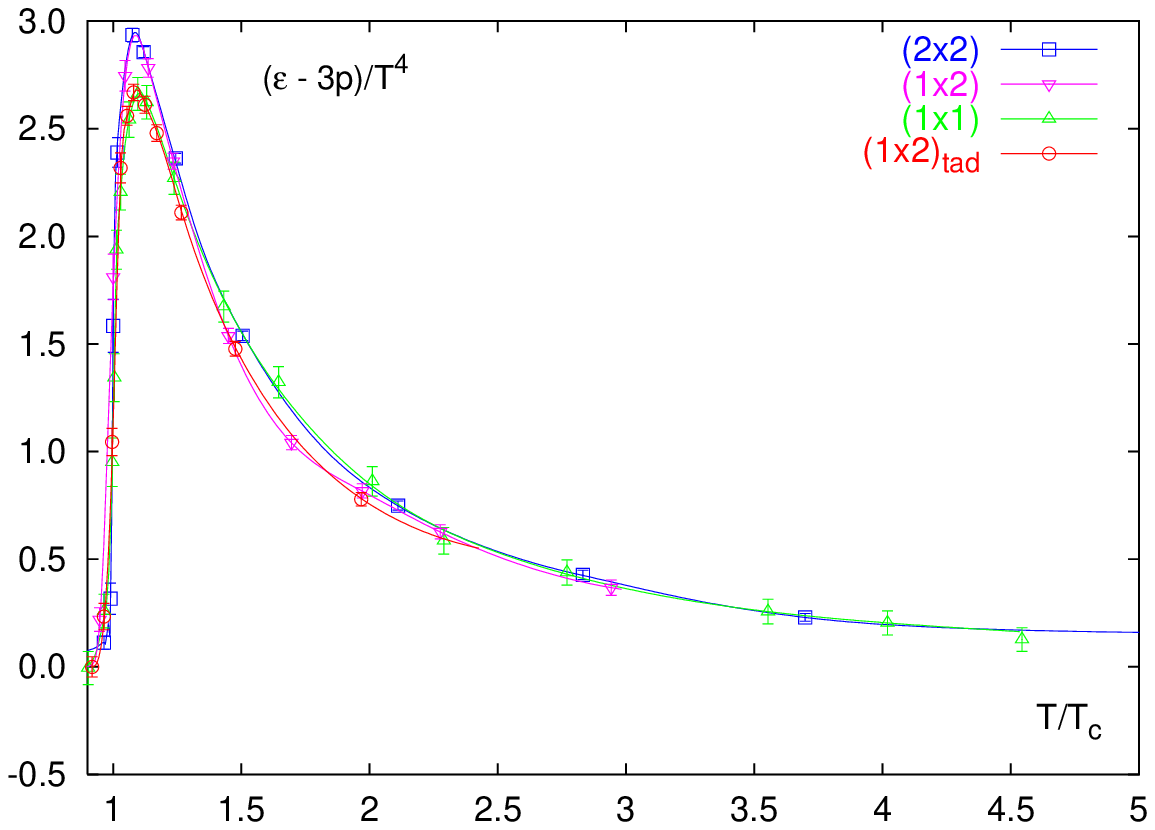,width=130mm}
\vspace*{-0.4cm}
  \epsfig{
       file=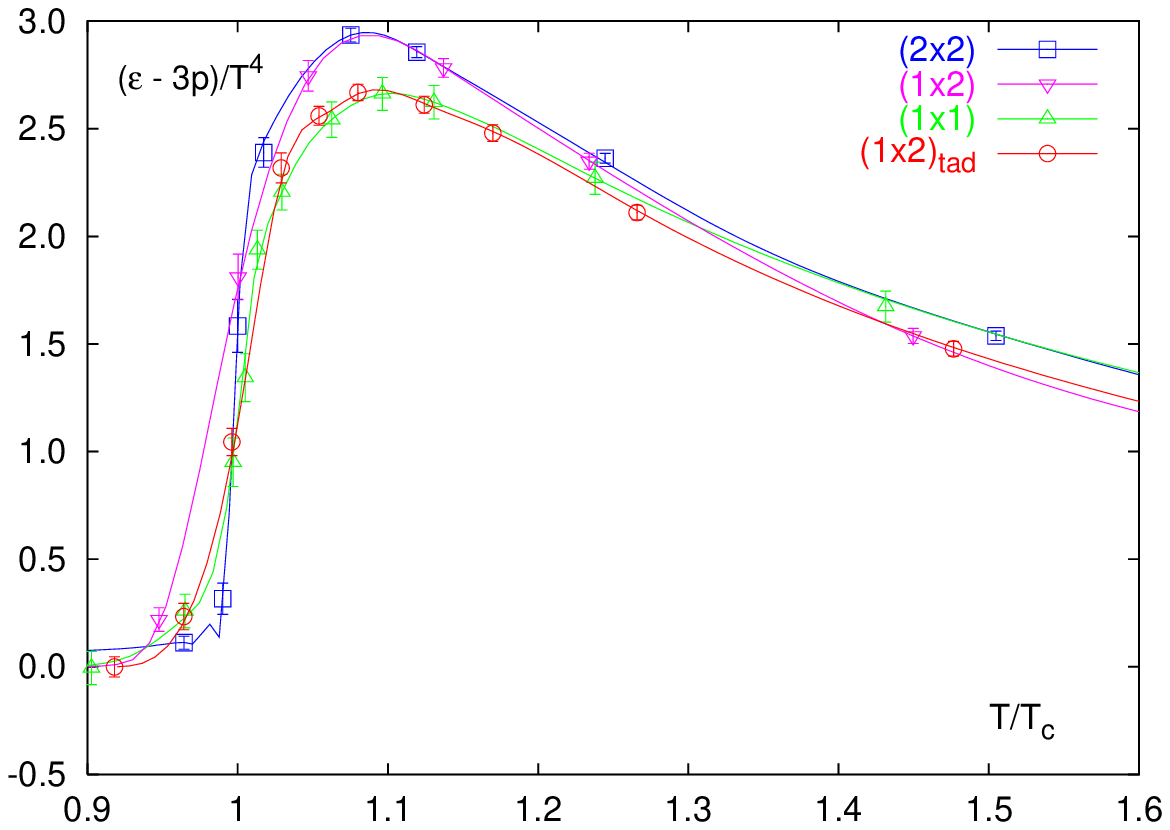,width=130mm}
\end{center}
\caption{
The difference $(\epsilon - 3p)/T^4$ versus $T/T_c$ calculated on lattices
with temporal extent $N_\tau=4$ for various improved actions. The results 
are compared to a calculation with the Wilson action on a $N_\tau=8$
lattice. In the lower figure the peak of $(\epsilon - 3p)/T^4$ is 
shown in detail.}
\label{fig:e3p}
\end{figure}
Using the temperature scale defined by the string tension calculations we 
also can extract the $\beta$-function, $a{\rm d} \beta / {\rm d} a$, outside 
the validity regime of the asymptotic 2-loop form, \ie 
in the coupling range explored here. With this, 
further thermodynamic observables can be extracted. In particular, we obtain 
for the difference between the energy density, $\epsilon$, and $3p$, 
\beqn
{\epsilon - 3p \over T^4} = \biggl({N_\tau \over N_\sigma}\biggr)^3 
\biggl(a{d \beta \over da}\biggr) \biggl[ \langle \tilde{S}\rangle_0- 
\langle \tilde{S}\rangle_T \biggr]~~.
\label{e3p}
\eqn

In Figure~\ref{fig:e3p} we compare the result obtained with tree level and
tadpole improved actions on lattices with temporal extent $N_\tau=4$ with 
corresponding results obtained with the Wilson action for $N_\tau=8$.
In the case of the Wilson action it has actually been observed that results
for $N_\tau=6$ and 8 coincide within errors and may thus be taken as 
the continuum limit result \cite{Boy96}.  
The good agreement we find here with the Wilson action
calculation confirms this observation, \ie cut-off effects are indeed small in
$(\epsilon -3p)/T^4$. In fact, the strong cut-off effects present in  the
ideal gas limit cancel exactly in this quantity, which in the high
temperature limit is ${\cal O} (g^4(T))$.  
The good agreement between results obtained with different actions also is an 
excellent consistency check for the determination of the temperature scales
and the analysis performed here.

\section{Conclusions}

We have analyzed thermodynamic properties of the SU(3) gauge theory using
tree level and tadpole improved actions. In general we find that the use 
of improved actions does lead to a significant reduction of the cut-off
dependence in observables like the energy density, the pressure and even the
surface tension at $T_c$. The improvement has, however, little effect on the
calculation of long-distance quantities like $T_c/\sqrt{\sigma}$. On the other
hand we do observe an improved scaling behaviour for the tadpole improved
action. There also might be a slight advantage in the use
of the tadpole improved action for the analysis of bulk thermodynamic 
observables like the pressure. However, at present this cannot be further 
quantified within the accuracy of our calculations.

It now seems that the systematic cut-off dependencies in calculations of
thermodynamic observables ($T_c$, equation of state, latent heat, surface
tension,..) are well under control with presently used improved actions.
Remaining ultra-violet cut-off dependences, which without doubt still are 
present, are hidden by statistical errors and/or inaccuracies in the 
determination of the observables. The latter result, for instance, also from 
infra-red cut-off
effects. This is true also for the determination of the string tension, 
which is
sensitive to the specific form of the fit used to analyze the heavy quark
potential at distance $R \simeq (0.25-1)$~fm. Such ambiguities are likely
to be the origin for the currently existing discrepancy between the
calculations presented here and in \cite{Iwa97}. 
However, also the thermodynamic calculations are still sensitive to
infra-red effects. While the finite volume effects are quite well under 
control for the determination of the critical temperature they certainly
have to be analyzed in more detail for the discontinuities (surface tension,
latent heat) at $T_c$.

\noindent
{\bf Acknowledgments:} The work presented here has been supported
through the Deutsche Forschungsgemeinschaft (DFG) under grant Pe 340/3-3.
The numerical work has been performed on the Quadrics Q4 and QH2 computers 
at the University of Bielefeld which have been funded by the DFG under grant  
Pe 340/6-1 .

\vfill\eject

\end{document}